\documentstyle[11pt]{article} 
\begin{document} 

\pagestyle{empty} 
\rightline{ULB-TH-99/06} 
\vspace{3.5cm} 
\begin{center} 
\LARGE{Gauge dilution and leptogenesis\\[20mm]} 
\large{S. Carlier, J.-M. Fr\`ere, F.-S. Ling\footnote{Aspirant F.N.R.S.}\\[8mm]} 
\small{Service de Physique Th\'eorique, CP225\\[-0.3em] 
                Universit\'e Libre de Bruxelles\\[-0.3em] 
Bvd du Triomphe, 1050 Brussels, Belgium \\[7mm]}

\vspace{1.5cm} 

\small{\bf Abstract}  
\\[7mm] 
\end{center} 
\begin{center} 

\begin{minipage}[h]{10.0cm} 
In this paper, we examine how gauge interactions can dilute the lepton asymmetry in lepton induced baryogenesis. Constraints 
imposed on Majorana masses keep this dilution at an acceptable level. 
\end{minipage}
 
\end{center} 

\newpage 

\pagestyle{plain} 

The out-of-equilibrium decay of heavy Majorana neutrinos can produce a net lepton 
asymmetry which will be later converted into a baryon asymmetry (see e.g.~\cite{Yanagida}). This simple scenario can explain 
today's observed value of asymmetry between matter and antimatter, meaned by the ratio ${{n_B}\over {s}}\simeq 10^{-10}$ with 
$n_B$ the net baryon density and $s$ the entropy. 

$L$-violating processes involving Majorana particles are only possible beyond the Standard Model (SM). In the minimal scheme, 
for each generation, one heavy right-handed neutrino with a Majorana mass $M$ is added to the SM. This neutrino has 
interactions with the other leptons and the usual Higgs scalars through the following coupling

\begin{equation} 
\label{lagrangian} 
-{\cal L}_{Yuk}=\overline{\psi_L}\Phi g_l e_R + \overline{\psi_L}\tilde{\Phi}g_{\nu} \nu_R +
{1\over 2}\overline{\nu_{R}^{c}}M\nu_R + h.c. 
\end{equation} 

\noindent where $ \Phi=\left( 
\matrix{{\phi}^+ \cr  
{\phi}^0} \right) $ is a Standard Model doublet and ${\psi}_L=\left( \matrix{{\nu}_L \cr e_L} \right) $. It is convenient to 
choose a basis in which the Majorana mass matrix is real and diagonal, so that the mass eigenstates above the electroweak 
phase transition are simply the Majorana neutrinos $ N=\frac{\nu_R+\nu_{R}^{c}}{\sqrt{2}} $ with mass $M$, while the 
left-handed leptons remain massless. Below the electroweak transition, the Yukawa couplings are related to the fermion masses 
($d$ refers to the Dirac contribution to neutrino masses, $l$ to the charged leptons): 

\begin{equation} 
m_d={{g_{\nu}v_L}\over{\sqrt {2}}}; \hfill m_l={{g_l v_L}\over{\sqrt {2}}} 
\end{equation} 

\noindent where $\frac{v_L}{\sqrt{2}}$ is the vacuum expectation value (vev) of 
$\Phi$. However, only the charged lepton masses are directly measured, while the neutrinos masses are currently only 
constrained. Moreover, the light neutrino masses are not given by the Dirac mass $ m_d $, but are obtained via the see-saw 
mechanism~(\cite{seesaw}). The range of $ m_d $ is thus highly speculative.

The lagrangian~(\ref{lagrangian}) shows that the Majorana neutrinos can decay either in left-handed lepton and 
$\Phi^{\dagger}$ or in right-handed antilepton and $\Phi$. A lepton asymmetry  arises only when CP invariance is violated, 
which is the case for complex Yukawa couplings. This CP asymmetry arises at the one-loop level, since 
it is given by the interference between tree-level diagrams and one-loop diagrams~(\cite{Yanagida},\cite{oneloop}) 

\begin{equation} 
\label{asymmetry} 
\epsilon_i={1\over {8\pi v^2}}\sum_j{{Im[({m_d}^{\dagger} m_d)_{ij}]^2}\over{({m_d}^{\dagger} m_d)_{ii}}}F(x_{ij}) 
\end{equation} 

\noindent where $x_{ij}={{M_j^2} \over {M_i^2}}$ and the function $F$ is given by 

$$ 
F(x)=\sqrt{x}[1-(1+x)\log{\mid {{1+x}\over{x}}\mid}] 
$$ 

\noindent The above formula only takes the vertex corrections into account and doesn't include the self-energy contributions. 
While this is obviously incomplete~(\cite{self}), it however gives the correct order of magnitude for cases where the mass 
hierarchy is well-established. Only for near mass degeneracy patterns can some technical and theoretical problems have a 
large impact~(\cite{buch}).

A departure from thermal equilibrium is the last requirement for the build-up of the lepton asymmetry. Here, this departure 
is set by the universe expansion rate. In the most favorable case, when $ \Gamma_N << H $, i.e. the Majorana neutrinos decay 
slowly compared to the universe expansion rate, the maximum value reached by the lepton asymmetry is given by  

\begin{equation} 
\frac{n_L}{s} \sim \frac{\epsilon_i}{g_*} , 
\end{equation} 

\noindent where $g_*$ counts the number of relativistic degrees of freedom at the epoch of the decay. This requirement puts 
the Majorana mass scale rather high, around $ M \sim 10^{10} GeV $, and suppresses strongly the light neutrino masses. 

We now turn to the question of embedding this scenario into a globally gauged theory. The inclusion of heavy right-handed 
neutrinos becomes natural only if the SM is embedded in a globally gauged theory involving gauge bosons in the right sector. 
The minimal scheme therefore naturally extends to a grand unification  theory like $ SO(10) $. Here, we will restrict our 
analysis to its subgroup $SU(2)_R \otimes SU(2)_L \otimes U(1)_{B-L}$. In this model, the vev of a scalar triplet $\chi_R$ of 
the right sector gives the Majorana mass to the right-handed neutrinos, as well to the gauge bosons $W_R^{\pm}$, $Z'$ 
\footnote{In more complicated models, extra scalars may contribute to the $W_R^{\pm}$ or $Z'$ masses}. The left-right phase 
transition is supposed to occur at very high temperatures for the purposes of leptogenesis.

As the mass of the mentionned gauge bosons and of the heavy neutrinos rely on the same scale and stem from the same source, 
those new degrees of freedom cannot be ignored. While the possible effects of the $Z'$ have been analysed in lepton-Higgs 
scattering processes~(\cite{buch2}), the existence of the gauge boson $ W_R $ implies the following additional decay 
channel(s) 
$$ 
N \rightarrow {\rm light lepton} + W_R 
$$ 
$W_R$ can be on mass-shell if the Majorana mass is bigger than the mass of $ W_R $, otherwise it is a virtual state that 
couples to the charged quark currents of the right sector. These decay channels cannot produce a lepton asymmetry, and so 
will act as a possible dilution of the bare asymmetry.  They will be taken into account by adding a dilution factor $X_i$ to 
the  CP asymmetry estimated above~(\ref{asymCP})

\begin{eqnarray} 
\label{asymCP2} 
\epsilon_i={{{{1\over {64\pi^2}}\sum_j {{\Im m[(m_{d}^{\dagger} m_d)_{ij}]^2}\over {v_L^4}}F(x_{ij})}\over{{(m_{d}^{\dagger} 
m_d)_{ii}}\over{8\pi v_L^2}}(1+X_i)}} 
\end{eqnarray} 

\noindent where the total decay width of the Majorana neutrino is given by 
$$ 
\Gamma_{N_i}=\Gamma_{i0} (1+X_i) 
$$ 
and $ \Gamma_{i0}={{(m_d^{\dag}m_d)_{ii}}\over {8\pi v^2}} M_i $ is the decay width into the $L$-violating channel. 

We will assume that the neutrino Dirac mass matrix has a form analoguous to the lepton mass matrix. This is however a 
basis-dependent statement. We will assume that in the basis where $M$ is diagonal, the neutrino Dirac mass matrix is given by 
the following popular pattern, where independent values are similar to those for charged leptons
\begin{equation}
m_d=\left(\matrix{ 
0 & m_{12} & 0 \cr 
m_{21} & m_{22} & m_{23} \cr 
0 & m_{32} & m_{33} \cr
} \right) 
\end{equation} 
where 
$$ 
\left\{\matrix{ 
\vert m_{12} \vert = \vert m_{21} \vert \simeq \sqrt{m_e m_{\mu}} \cr 
\vert m_{22} \vert \simeq m_{\mu} \hfill \cr 
\vert m_{33} \vert \simeq m_{\tau} \hfill \cr 
\vert m_{23} \vert = \vert m_{32} \vert \simeq \sqrt{m_{\mu}m_{\tau}} \cr 
}\right.   
$$ 
This choice is a typical one for low energy mass matrices 
and is especially well-motivated in the case of a $LR$ symmetry (see below).
Using this matrix and assuming maximal phases, the estimate for the bare CP asymmetries become (in absence of any dilution)
\begin{equation} 
\label{asymCP} 
\left.\matrix{
\epsilon_1\simeq  [1.1; 3.3] \cdot 10^{-8} \cr 
\epsilon_2\simeq  [2; 6] \cdot 10^{-7}\hfill \cr
\epsilon_3\simeq [1.4; 4.2]\cdot 10^{-8} \cr
}\right.
\end{equation} 
where the allowed range reflects the possible values of the function $F(x)$ in~(\ref{asymCP}) for a wide range of mass 
hierarchies. We see that the bare asymmetry is the biggest for the second generation, and that the obtained estimates are 
compatible with the needed value of the asymmetry, since we expect $g_* \sim 10^2$ at the epoch of interest. For the first 
and the third generations, we see that the total dilution factor may not exceed $ \sim 1 $ to yield the observed value. For 
the second generation, a dilution factor of order $ \sim 10 $ remains possible.  

We will now examine the dilution channels in two stages :  

\begin{itemize} 
\item{when $M>M_{W_R}$: the dilution will mainly occur through the faster 2-body decay channel, which alone will be taken 
into 
account in the evaluation of the dilution factor.} 
\item{when $M<M_{W_R}$: only the 3-body decay channels will be present.} 
\end{itemize} 
First we consider the case $M>M_{W_R}$. The 2-body decay rate of the Majorana neutrino $ N_i $ is 
\begin{equation}
\label{2body}
\Gamma_{2i}=\frac{g_R^2}{32\pi} \frac{M_i^3}{M_{W_R}^2} (1- \frac{M_{W_R}^2}{M_i^2})(1+2 \frac{M_{W_R}^2}{M_i^2})
\end{equation}
\noindent and takes the two channels $N \rightarrow l_R + W_R^+$ and $N \rightarrow \overline{l}_R + W_R^-$ into account.
This gives, for the dilution factor 
$$ 
X_i=\frac{g_R^2}{2(g_{\nu}^{\dagger} g_{\nu})_{ii}} \frac{(1-\alpha_i)(1+2 \alpha_i)}{\alpha_i} , 
$$ 
where $ \alpha_i=\frac{M_{W_R}^2}{M_i^2}$. In the $LR$ symmetric models the gauge couplings for the right and the left 
sectors are similar, so that, for $\alpha = 1/2$, we deduce the following estimates 
\begin{equation} 
\label{2bodydil} 
\left.\matrix{
X_1 \sim 10^8 \cr
X_2 \sim 10^5 \cr
X_3 \sim 10^3 \cr
}\right.
\end{equation} 

\noindent As these values increase for lower $\alpha$, the dilution due to the 2-body decay channel ruins completely the 
produced asymmetry and is therefore not acceptable. The leptogenesis scenario can be realistic only if the lightest decaying 
Majorana neutrino is lighter than the charged right gauge boson. This result was previously obtained~(\cite{sarkar}), but the 
argument was only valid in a supersymmetric theory. 

We thus turn to the second case, $M<M_{W_R}$. The decay rate of the Majorana neutrino $N_i$ in the 3-body channel can be 
bound 
by the following lower value 
\begin{equation}
\Gamma_{3i} \geq  \frac{3g_R^4}{2^{11} \pi^3} \frac{M_i^5}{M_{W_R}^4}
\end{equation}
It has been obtained using a point-like interaction, and takes all possible quark channels into account. The corresponding 
dilution factor is 
$$ 
X_i = \frac{3}{8 \pi^2} \frac{\lambda_i^4}{(g_{\nu}^{\dagger} g_{\nu})_{ii}} , 
$$ 
where $ \lambda_i = \frac{M_i}{v_R} $ represents the coupling constant of the right-handed neutrino to the scalar triplet $ 
\chi_R $ above the left-right symmetry breaking phase transition. We see that the dilution constraint translates into an 
upper bound on $\lambda$. Finally, a lower bound can be put on the Majorana mass by the out-of-equilibrium condition. The 
table~(\ref{3bodydil}) gives typical values
\begin{equation}
\label{3bodydil}
\left.\matrix{
\lambda_1 < 7 \cdot 10^{-3} & M_1 > 10^{7}Gev \cr
\lambda_2 < 0.17 & M_2 > 3 \cdot 10^{11}Gev \cr
\lambda_3 < 0.11 & M_3 > 4 \cdot 10^{11}Gev \cr
}\right.
\end{equation}

\noindent The bounds for $\lambda$ show that the embedding of the leptogenesis secenario in a gauged theory imposes non 
negligible constraints on the Yukawa couplings. While these couplings can be very small in the SM, there is however no 
natural explanation for such a smallness, and values of order $\sim O(1)$ can be preferred. In these conditions, a 
leptogenesis scenario based on the decay of $N_2$ rather than $N_1$ seems more feasible. This implies that $M_2 < M_1$ and 
puts the leptogenesis scale around $10^{11} GeV$. Of course, a leptogenesis scenario based on the decay of $N_1$ remains 
possible if $\lambda_1$ is tuned to be sufficiently small. In this case, the leptogenesis can occur at a lower energy scale, 
but this is done at the cost of fine-tuning.

\newpage 
\vskip 1\baselineskip
\noindent {\large \bf Conclusion}
\vskip 1\baselineskip

The Majorana neutrino leptogenesis scenario is severely constrained when the heavy Majorana neutrinos are embedded in a gauge 
theory. The lightest Majorana neutrino must be lighter than the charged gauge bosons of the right sector to prevent excessive 
dilution. Moreover, chosing a usual pattern for the Yukawa couplings at low energy, we see that the lightest Majorana 
neutrino must be considerably lighter than the gauge boson to make their effects negligible. Finally, a larger dilution 
factor is allowed for the Majorana neutrino of the second flavour. This leads, through the see-saw mechanism, to a 
suppression of the corresponding light neutrino mass with respect to the first generation.

\vskip 1\baselineskip
\noindent {\large \bf Acknowledgements}
\vskip 1\baselineskip

We thank M. Tytgat and V. Van Elewyck for helpful discussions. This work was partially supported by the IISN (Belgium) and by 
the Communaut\'e Fran\c caise de Belgique - Direction de la Recherche Scientifique programme ARC.

\end{document}